\documentclass[aps,prb,twocolumn,groupedaddress]{revtex4}
\usepackage{graphicx}

\begin{document}

\title{Electrical Control of Optical Properties of Monolayer MoS$_2$}

\author{A.K.M. Newaz$^1$, D. Prasai$^2$, J.I. Ziegler$^1$, D. Caudel$^{1,3}$, S. Robinson$^4$, R.F. Haglund Jr.$^{1,2}$ and K.I. Bolotin$^{1,2}$}
\address{$^1$Department of Physics and Astronomy, Vanderbilt University, Nashville, Tennessee 37235-1807, USA}
\address{$^2$Interdisciplinary Graduate Program in Materials Science, Vanderbilt University, Nashville, Tennessee 37234-0106, USA}
\address{$^3$Department of Physics, Fisk University, Nashville, Tennessee 37208, USA}
\address{$^4$Department of Chemistry and Physics, Belmont University, Nashville, Tennessee 37212, USA}
\email[e-mail address: ]{kirill.bolotin@vanderbilt.edu}

\date{\today}

\begin{abstract}

We investigate electrical gating of photoluminescence and optical absorption in monolayer molybdenum disulfide (MoS$_2$) configured in field effect transistor geometry. We observe an hundredfold increase in photoluminescence intensity and an increase in absorption at $\sim 660$ nm in these devices when an external gate voltage is decreased from +50 V to -50 V, while the photoluminescence wavelength remains nearly constant. In contrast, in bilayer MoS$_2$ devices we observe almost no changes in photoluminescence with gate voltage. We propose that the differing responses of the monolayer and bilayer devices are related to the interaction of the excitons in MoS$_2$ with charge carriers.

\end{abstract}

\maketitle


Materials with electrically controllable optical properties find uses in diverse applications ranging from electro-optical modulators to display screens. Unfortunately, the optical constants of most bulk semiconducting materials do not vary significantly with electric field. In the case of silicon, for instance, the variation in refractive index with gate voltage is smaller than $0.01\%$, limiting the footprint and the modulation depth of electro-optical modulators\cite{Liu04}. While larger electro-optical response has been demonstrated in other semiconductors, such as germanium and gallium arsenide, integration of these materials with silicon CMOS fabrication may prove difficult\cite{Miller84,Kuo05}. Very recently, two-dimensional (2D) atomic crystals\cite{Novo05} emerged as a potential alternative to bulk semiconductors for photonic applications\cite{Bona10}. In graphene, the most widely studied 2D material, changes in optical absorption larger than $100\%$ produced by the electric field effect have been used to demonstrate nanoscale electro-optical modulators in the infrared range\cite{Liu11}. However, the lack of a band gap in graphene makes its uses at visible frequencies infeasible.

\begin{figure}[thp!]
 \centering
\includegraphics[width=3.4 in,angle=0,clip=]{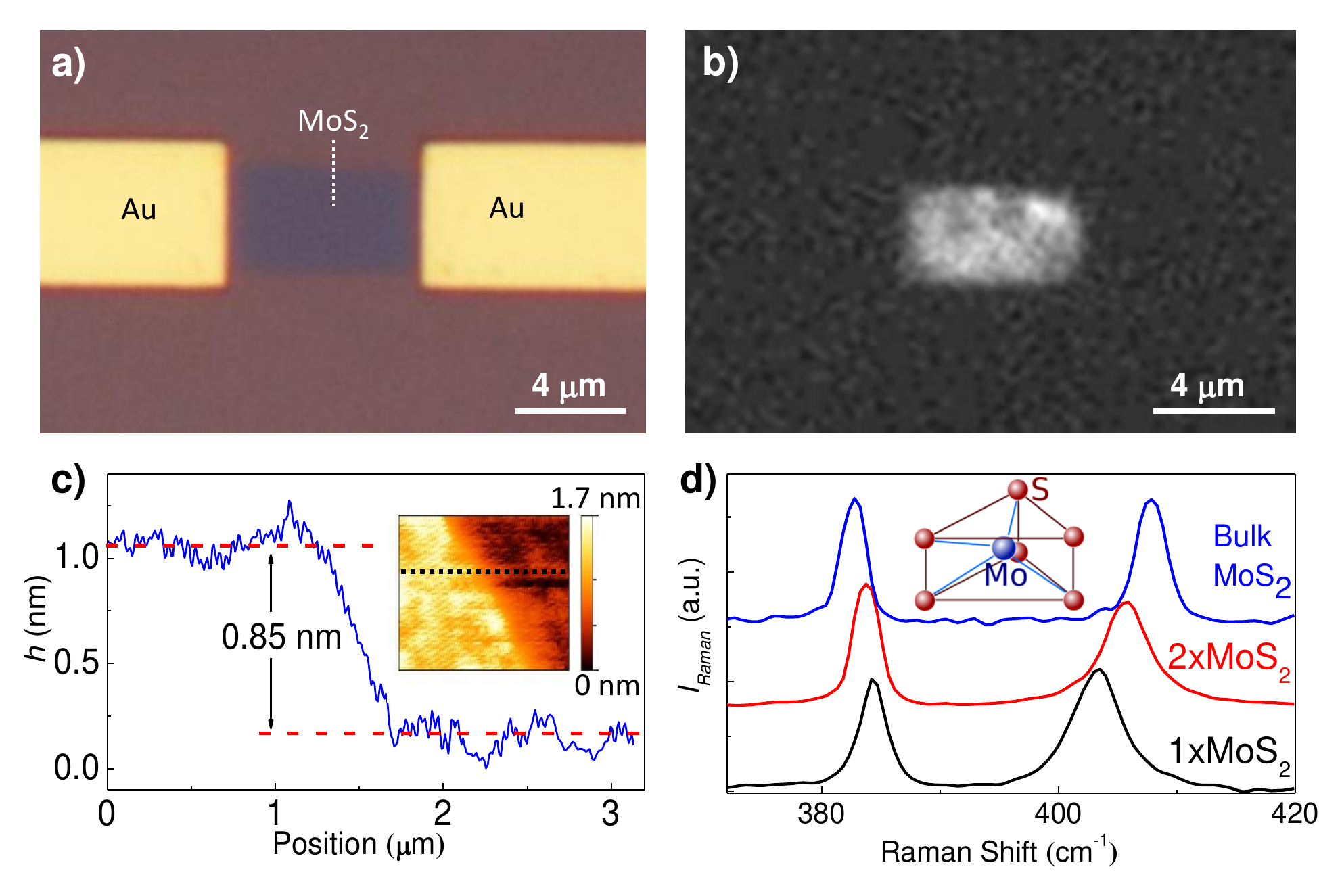}
\caption[Fabrication of monolayer MoS$_2$ field effect  transistors] {\label{fig:fig1} Fabrication of monolayer MoS$_2$ field effect transistors. a) Optical image of an electrically contacted MoS$_2$ on top of SiO$_2$/Si substrate (Device $\#1$); the gold electrodes are deposited after the MoS$_2$ flake is exfoliated onto the substrate. b) Fluorescence image of the same device under green light (530-590nm) excitation. Bright fluorescence is a signature of monolayer MoS$_2$. c) Contact-mode atomic force microscopy of the same device confirms monolayer nature of MoS$_2$. d) Raman spectroscopy data for single-, bi- layer MoS$_2$ and bulk MoS$_2$ specimens. The inset shows trigonal prismatic structure of MoS$_2$ unit cell.}
\end{figure}

Here we demonstrate electrical control of photoluminescence quantum yield and absorption coefficient in the visible range for a different two-dimensional crystal, monolayer molybdenum disulfide (MoS$_2$). This material consists of a layer of molybdenum atoms surrounded by sulfur in a trigonal prismatic arrangement\cite{Yoffe73}  (Fig. 1d, Inset).  Unlike semi-metallic graphene, monolayer MoS$_2$ (1xMoS$_2$) is a semiconductor with a direct band gap of $\sim 1.85$ eV, and is therefore optically active in the visible range\cite{Mak10,Splen10}. The combination of a substantial band gap and high ($> 200$ cm$^2$/V$\cdot$s) carrier mobility \cite{Radi11} invites electro-optic applications of MoS$_2$. Finally, monolayer MoS$_2$ can be synthesized using several scalable methods potentially compatible with standard CMOS technology\cite{Lee12}. We fabricate monolayer MoS$_2$ field effect transistors (FETs) and probe changes in their optical properties in response to an externally applied gate voltage ($V_G$). At $V_G$=-50 V, we observe a bright photoluminescence (PL) band centered at $\sim 1.85$ eV that decreases in intensity by more than a factor of 100 as $V_G$ is swept from -50 V to 50 V. Concurrently, we observe a decrease in absorption  at the same wavelength. We propose that these phenomena are caused by the interaction of excitons in MoS$_2$ with conduction electrons via the phase-space filling effect.

We have fabricated FETs by first depositing monolayer and bilayer MoS$_2$ flakes with average dimensions of several microns onto SiO$_2$/Si substrates via micromechanical exfoliation\cite{Novo05}. Individual MoS$_2$ flakes are contacted electrically using metal electrodes deposited via electron beam lithography followed by thermal metal evaporation (Fig. 1a). While Cr/Au and Ti/Au contacts were used, we found that Au electrodes without any wetting layer produced the least contact resistance \cite{Pop12}. Altogether we fabricated eight monolayer MoS$_2$ (1xMoS$_2$) FETs that showed similar electrical and optical characteristics. We also fabricated two bilayer MoS$_2$ (2xMoS$_2$) devices. We confirmed the monolayer character of the 1xMoS$_2$ samples in three different ways. First, we investigated the fluorescence microscopy images of the devices, since bright fluorescence only occurs for monolayer MoS$_2$, and is a signature of a direct band-gap material (Fig. 1b)\cite{Mak10,Splen10}. Second, contact-mode atomic-force microscopy (AFM) measurements confirm that our devices are less than 1 nm thick, comparing favorably to the expected value of $\sim 0.7\,$nm (Fig. 1c) \cite{Lee10}. Finally, Raman spectra of our devices exhibit characteristic $A_{1g}$ and $E_{2g}$ peaks that are spaced 19 cm$^{-1}$ apart, a characteristic signature of monolayer MoS$_2$ flakes (Fig. 1d)\cite{Lee10}.   The bilayer character of MoS$_2$ in the 2xMoS$_2$ devices was also confirmed by Raman spectroscopy, fluorescence microscopy and AFM.

We first measured electrical transport in a typical 1xMoS$_2$ device $\#1$ in ambient dark environment at room temperature. The source-drain current-voltage curve, $I_{sd}(V_{sd})$, remains linear for $|V_{sd}|<50$ mV (Fig. 2a, Inset), indicating good electrical contact to the device and the lack of any Schottky barriers at the electrode-MoS$_2$ interface \cite{Radi11}. The conductance $dI_{sd}/dV_{sd}$ in this device can be controlled by an externally applied gate voltage $V_G$, due to the electric-field effect in MoS$_2$ (Fig. 2a). The conductivity increase for $V_G>0$ V proves that the device is operating in the electron-doping regime, while near-complete suppression of conductivity for $V_G <0$ V is consistent with the larger than $k_B T$ band gap of MoS$_2$ \cite{Radi11}. For device $\#1$, we estimate the field-effect mobility $\mu_{FE}=(L/WC_G)(dR^{-1}/dV_G)$ cm$^2$/V$\cdot$s, where $L, W$ and $R$ are the length, width and resistance of the device, respectively and $C_G\sim 116$ aF $\mu$m$^{-2}$ is the geometrical capacitance between MoS$_2$ and the silicon back gate. For other devices, we observed mobilities in the range $0.3-60$ cm$^2$/V$\cdot$s.

\begin{figure}[thp!]
 \centering
\includegraphics[width=3.4in,angle=0,clip=]{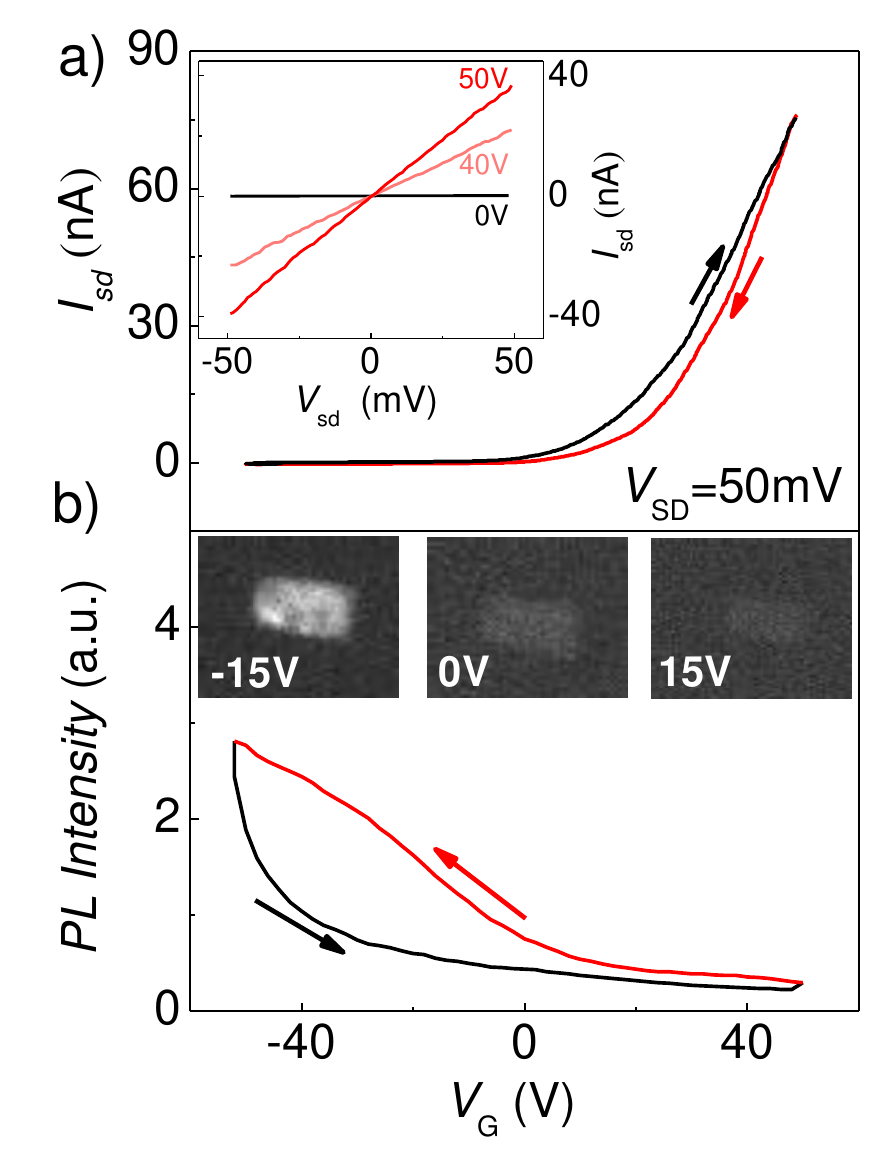}
\caption[Electrical and optical characterization of monolayer MoS$_2$ FETs.] {\label{fig:fig2} Electrical and optical characterization of monolayer MoS$_2$ FETs. a) Source-drain current $I_{sd}$ vs. gate voltage $V_G$ at applied source drain bias $V_{sd}=50$ mV for Device $\#1$ measured in the dark (Also see supplementary Fig. S4). Arrows are indicating the direction of $V_G$ sweeping: from -50 V to +50 V and then back to -50 V.  The inset demonstrates the linearity of $I_{sd} (V_{sd})$ curves for -50 mV$<V_{sd}<$50 mV. b) The integrated intensity of photoluminescence vs. $V_G$ for the same device. The excitation wavelength was 2.33 eV, power $\sim 1\,\mu$W, and the beam spot size $\sim 1\,\mu$m. The inset shows fluorescence microscopy images of the same device at three different gate voltages.}
\end{figure}

Simultaneously with electrical measurements, we studied photoluminescence of the devices, both via conventional fluorescence microscopy, and by using scanning confocal microscopy with laser excitation wavelength at $\sim 532$ nm (2.33 eV), power ranging between $1-200$ $\mu$W, and with a diffraction-limited spot size of $\sim 1\,\mu$m. At zero gate voltage we observe bright luminescence at $\sim 1.85$ eV (feature ``A'', Fig. 3a), a feature previously observed both in monolayer \cite{Mak10,Splen10,Eda11} and bulk \cite{Yoffe73} MoS$_2$. This feature has been attributed to the recombination of photoexcited excitons across the direct band gap at the K-point (Fig. 3a, Inset).

\begin{figure}[thp!]
 \centering
\includegraphics[width=3.4 in,angle=0,clip=]{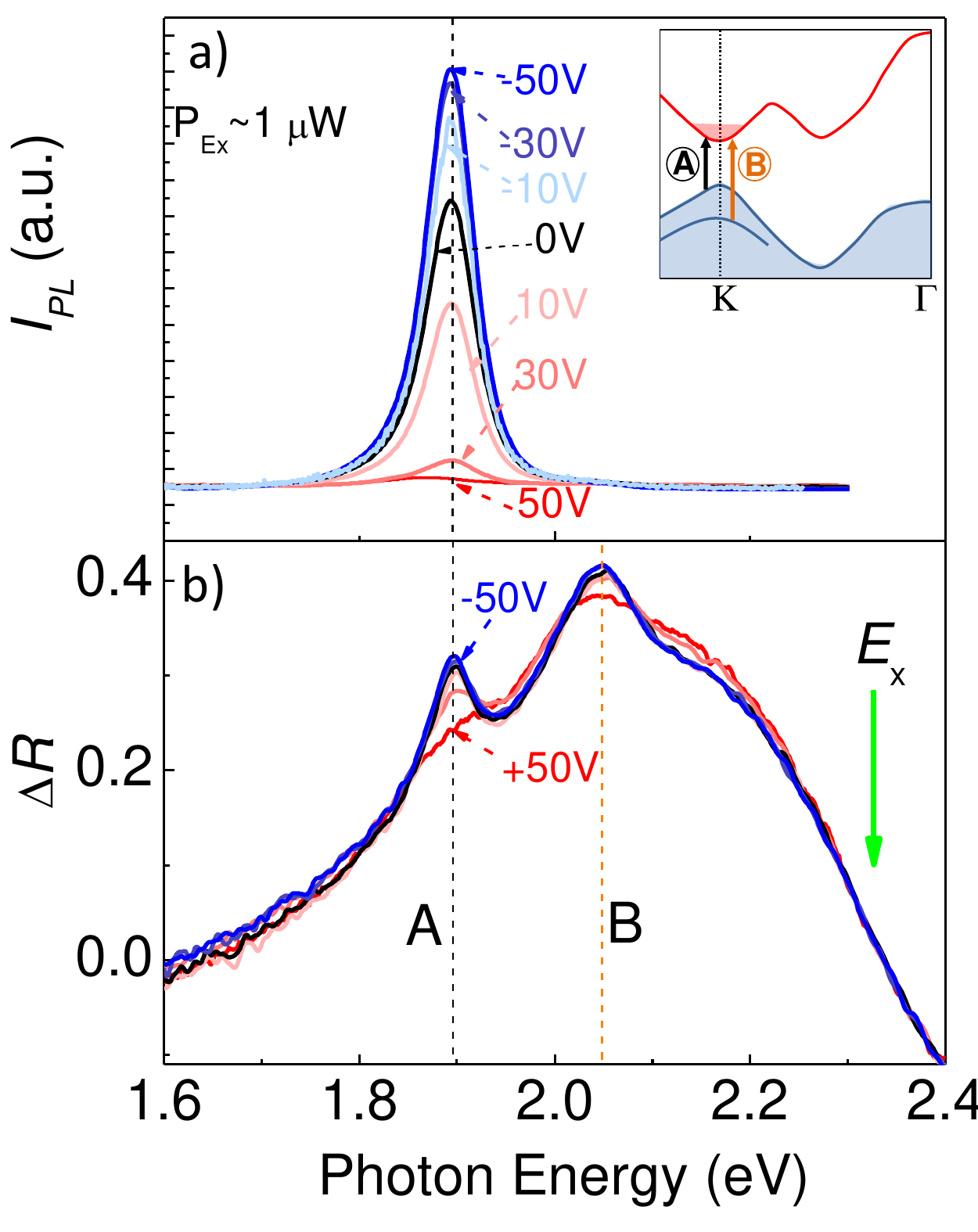}
\caption[ Photoluminescence and absorption spectra in monolayer MoS$_2$ (Device $\#2$). ] {\label{fig:fig3} Photoluminescence and absorption spectra in monolayer MoS$_2$ (Device $\#2$). a) PL spectra taken at different gate voltages ($V_G$=-50V, -30V, -10V, 0V, 10V, 30V and 50V) at low excitation power ($1\,\mu$W). The inset shows the band diagram of monolayer MoS$_2$; direct band-gap exciton transitions ``A'' and ``B'' are indicated by arrows. b) Differential reflectivity ($\Delta R$) spectra for the same device in the same range of gate voltages. Low-power white-light illumination was used. The green arrow indicates the position of the laser excitation energy (2.33 eV) used to record PL data in a).}
\end{figure}

Crucially, in every device, the PL intensity changes dramatically with gate voltage. When $V_G$ is increased, with a concomitant increase in conductivity, the intensity of the PL (integrated area under the peak) diminishes (Fig. 2b). In the range of gate voltages between +50 V and -50 V, the maximum PL intensity changes by more than factor of 12 for that device. Gate-dependent variation in the PL intensity up to $\sim 160$ has been observed for other devices, such as device $\#2$ (Fig. 3a). This variation was found to be fully reversible, reproducible over months of measurements, and persistent in the entire range of excitation powers (Supplementary Information, Fig. S2).

Next, we focus on the absorption coefficient $\alpha(h\nu)$  of monolayer MoS$_2$ and investigate its possible dependence on the gate voltage $V_G$. We accomplish this by measuring differential reflectivity $\Delta R$ of our devices, where $\Delta R\equiv(R_{\textrm{off}}-R_{\textrm{on}})/R_{\textrm{off}},$ and $R_{\textrm{on}}(R_{\textrm{off}})$ is the reflectivity of the MoS$_2$ specimen on SiO$_2$/Si substrate (bare substrate next to MoS$_2$). We observe a prominent peak in $\Delta R (h\nu)$ at an energy corresponding to the feature ``A'' in the PL spectrum and an additional peak ``B'' at $\sim 2$ eV (Fig. 3b). These features correspond to excitonic transitions between the valence band split by spin-orbit interaction and the conductance band (Fig. 3a, Inset) \cite{Splen10}. Crucially, we find that while both ``A'' and ``B'' peaks in $\Delta R$ depend on gate voltage, away from these peaks $\Delta R$ is $V_G$-independent. Since $\alpha(h\nu)$ and $\Delta R$ are interrelated \cite{Fork12}, we conclude that absorption is constant away from ``A'' and ``B'' peaks.

\begin{figure}[ht]
 \centering
\includegraphics[width=3.0in,angle=0,clip=]{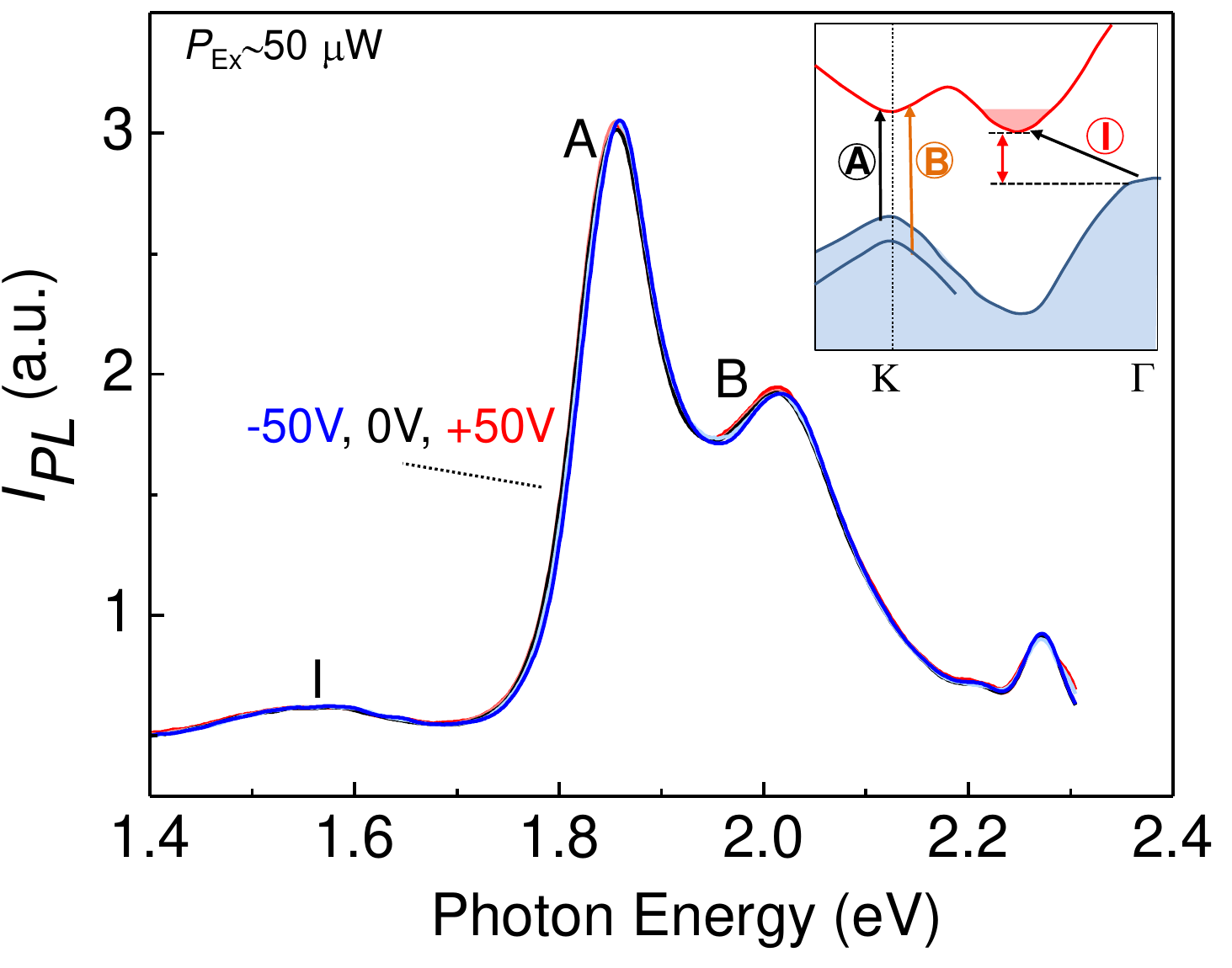}
\caption[Photoluminescence of bilayer MoS$_2$. ] {\label{fig:fig4} Photoluminescence of bilayer MoS$_2$. PL spectra of a bilayer MoS$_2$ device recorded at gate voltages -50V,-30V, 0V, 30V, 50V under 50 $\mu$W laser excitation power. Spectra at different $V_G$ are very similar and collapse onto the same curve. The inset shows the band structure of bilayer MoS$_2$. Along with ``A'' and ``B'' transitions, momentum-violating transition ``I'' across indirect band gap of bilayer MoS$_2$ is indicated.  }
\end{figure}

Finally, we investigate the $V_G$-dependence of PL for a bilayer MoS$_2$, an indirect band gap material \cite{Mak10,Splen10}.  For $V_G=0$ V, we observe features ``A'' and ``B'', similar to those seen in single-layer MoS$_2$, along with an appearance of a broad lower-energy feature ``I'' at $\sim 1.6$ eV (Fig. 4). We use larger excitation power $\sim 50 \mu$W since the overall luminescence yield in these devices is significantly lower than that from monolayer MoS$_2$ \cite{Splen10}. (Data taken using smaller excitation power are presented in Supplementary Figure S5). These spectral features are related to the band structure of 2xMoS$_2$ (Fig. 4, Inset). The low-intensity feature ``I'' is associated with momentum-violating phonon-assisted transition across the indirect band gap\cite{Mak10,Splen10}. Calculations predict that the band structure near the K-point – and hence excitons ``A'' and ``B'' – is only weakly affected by quantum confinement and is similar between single- and bilayer MoS$_2$. The comparable intensities of both features ``A'' and ``B'' in 2xMoS$_2$ are also unsurprising, since in 2xMoS$_2$ both of these transitions result from hot luminescence. Crucially, for 2xMoS$_2$ we observe no changes larger than $\sim 7\%$ in either feature ``A'' and ``B'' in the accessible range of $V_G$ (Fig. 4 and Fig. S5).

Summarizing the discussion so far, our main experimental findings are i) large variation of both PL intensity and optical absorption with gate voltage for monolayer MoS$_2$, and ii) the lack of substantial PL variation for bilayer MoS$_2$. We now focus on elucidating the mechanism of these phenomena.

First, the observed changes in PL intensity are not a result of electroluminescence \cite{Nobo11}. Measured photoluminescence is relatively constant across the devices' area, does not depend on the bias voltage applied to MoS$_2$, and was observed for zero bias current. Second, the observed changes in PL intensity are not caused by changes in absorbance of MoS$_2$ at the excitation frequency. Indeed, since $\Delta R(h\nu)$ does not vary with $V_G$ at the excitation energy $h\nu=2.33$ eV (Fig. 3b), away from ``A'' and ``B'' peaks, absorption coefficient $\alpha(h\nu=2.33$ eV) must be independent of gate voltage. This result is expected: in the measured gate voltage range $\Delta V_G=100$ V, the expected change in the carrier density due to the field effect is $\Delta n=C_G\Delta V_G/e\sim 7\times 10^{12}$ cm$^{-2}$. This change in the carrier density translates into a shift of the Fermi energy by  $\pi\hbar^2 n/m_e \sim 60$meV, where m$_e\sim 0.3m_0$  is the effective electron mass in 1xMoS$_2$ ( assuming spin degeneracy for the conduction band)\cite{Chei12}. This shift is small compared to the difference between fluorescence (1.85 eV) and excitation (2.33 eV) energies. Therefore, an electrostatically induced shift of the Fermi energy cannot affect the absorption at the excitation wavelength, opposite to what is observed for graphene in the infrared range\cite{Wang08}.

Next, we consider the possible contribution of sample disorder. In principle, defects and disorder in a material can localize both charge carriers and excitons, which, in turn, can result in gate-voltage-dependent photoluminescence \cite{Fink95}. To analyze this scenario, we compared PL data from different samples with mobility ranging from 0.1 cm$^2$/Vs to 13 cm$^2$/Vs. Despite over two orders of magnitude variation in carrier mobility between samples, $I_{PL}(V_G)$ curves were similar for every measured device, with less than a factor of two variation of PL intensity recorded at $V_G=-50$ V between different samples (see Supplementary Material, Fig. S3). While we did observe larger variation of PL intensity between devices at $V_G=+50$ V (such as Devices $\#1$ and $\#2$), this variation can be ascribed to difference in unintentional doping levels between the samples resulting from interactions with the substrate \cite{Shi09}. We therefore believe that the mechanism responsible for gate-voltage-dependent PL intensity is intrinsic rather than extrinsic in nature and is primarily related to the interaction of excitons in MoS$_2$ with free charge carriers.

We now suggest a possible mechanism for this interaction, the phase-space filling effect \cite{Sch85}.  In this mechanism, an increase of the carrier density renders part of a phase space unavailable for exciton formation due to Pauli exclusion principle. This causes a reduction in the exciton oscillator strength and a corresponding decrease of PL intensity and excitonic absorption.  A simple estimate \cite{Sch85} predicts that the PL intensity will be halved at the critical carrier density $n=2/\pi a_0^2 \sim 6\times 10^{13}$ cm$^{-2}$, where $a_0\sim 1$ nm is an effective Bohr radius for an exciton \cite{Chei12}. While this density is an order of magnitude larger than the variation of the carrier density $\Delta n\sim 7\times 10^{12}$ cm$^{-2}$ in our experiment, it is possible that this deviation is caused either by inaccuracies in the estimated exciton radius stemming from uncertainty in the dielectric constant of MoS$_2$ (for a monolayer MoS$_2$, the effective dielectric constant could be affected by either the underlying substrate or by the impurities on the surface of MoS$_2$) or by effects related to nonuniform doping profiles in the devices.  Furthermore, the phase-space filling mechanism is consistent with the absence of gate-dependent changes in PL in bilayer MoS$_2$. Indeed, for 2xMoS$_2$, the excitons and the conductions electrons occupy different regions of the phase space. The one-particle states participating in the formation of ``A'' and ``B'' excitons have momenta near the K-point, whereas conduction electrons reside across the indirect gap, away from the K-point  (Fig. 4, Inset). Therefore, changes in the carrier density should not affect the excitonic absorption and PL intensity for 2xMoS$_2$.

In summary, we have demonstrated that both photoluminescence and absorption of monolayer MoS$_2$ at $\sim 1.85$ eV can be controlled by gate voltage. We propose that this effect in MoS$_2$ is due to the interaction of excitons with charge carriers and suggest a possible mechanism for such an interaction through the phase-space filling effect. We expect that time-dependent PL measurements, as well as measurements at cryogenic temperatures will elucidate the origin of the observed phenomena \cite{Korn11}.

We envision multiple potential applications for monolayer devices of this type. First, the optical readout of the electronic states of MoS$_2$ transistors can be employed to investigate the nature of conduction in this material and to realize various optoelectronic devices. Second, electrically controlled absorption of light and photoluminescence in high-mobility MoS$_2$ can be utilized to create nanoscale electro-optical modulators operating in the visible range. Finally, we envision the possibility of controlling absorption and fluorescence wavelength in similar devices by exploiting other monolayer materials from the dichalcogenide family, such as MoSe$_2$, WS$_2$ and many others \cite{Yoffe73}. 

{\bf{Acknowledgements:}}  KIB acknowledges the support through NSF CAREER DMR-1056859 and NSF EPS 1004083. JIZ and RFH acknowledge support from the Defense Threat Reduction Agency (HDTRA1-1-10-0047). Samples for this work were prepared at the Vanderbilt Institute of Nanoscale Science and Engineering using facilities renovated under NSF ARI-R2 DMR-0963361. We gratefully acknowledge Kirill A. Velizhanin, Sokrates Pantelides, Leonard Feldman, Arend van der Zande for useful discussions and Hiram Conley for technical assistance.

\end{document}